# Orientation-dependent superconductivity and electronic structure of the rare-earth metal/$KTaO_3$ interfaces

Guowei Yang,[1#*] Weifan Zhu,[1#] Jiawen Zhang,[1] Hao Zheng,[1] Yi Wu,[1] Huali Zhang,[1] Ge Ye,[1] Dajun Su,[1] Yanan Zhang,[1] Chao Cao,[1] Xin Lu,[1] Huiqiu Yuan,[1] and Yang Liu[1,2*]

[1]Center for Correlated Matter and Department of Physics, Zhejiang University, Hangzhou 310058, China

[2]Collaborative Innovation Center of Advanced Microstructures, Nanjing University, Nanjing 210093, China

#These authors contribute equally to the work

*Corresponding author. Email: gwyang@zju.edu.cn

*Corresponding author. Email: yangliuphys@zju.edu.cn

## Abstract

The recent discovery of orientation-dependent superconductivity in $KTaO_3$-based interfaces has attracted considerable interest, while the underlying origin remains an open question. Here we report a different approach to tune the interfacial electron gas and superconductivity by forming interfaces between rare-earth (RE) metals (RE being La, Ce, Eu) and $KTaO_3$ substrates with different orientations. We found that the interfacial superconductivity is strongest for the Eu/$KTaO_3$ interfaces, becomes weaker in La/$KTaO_3$ and is absent in Ce/$KTaO_3$. Using *in-situ* photoemission, we observed distinct valence bands associated with RE metals, as well as a pronounced orientation dependence in the interfacial electronic structure, which can be linked to the orientation-dependent superconductivity. The photoemission spectra show similar double-peak structures for the (111) and (110) oriented interfaces, with an energy separation close to the LO4 phonon of $KTaO_3$. Detailed analyses suggest that this double-peak structure could be attributed to electron-phonon coupling, which might be relevant for the interfacial superconductivity.

# I. INTRODUCTION

A variety of interesting physical phenomena, which do not exist in bulk forms, can emerge at the oxide interfaces [1,2]. Interfacial superconductivity and magnetism in high-mobility electron gas between $LaAlO_3/SrTiO_3$ is a classic example [3,4,5]. The recent discovery of superconductivity in $KTaO_3$-based interfaces with much higher superconducting critical temperature ($T_c$) and large orientation dependence has attracted widespread attention [6,7,8]. Both $EuO/KTaO_3$ and $LaAlO_3/KTaO_3$ interfaces share similar superconductivity properties, whose maximum $T_c$ is ~2 K for (111) interfaces, decreases to ~1 K for (110) interfaces, and becomes non-superconducting for (001) interfaces. Such an anisotropic behavior is unique for this interfacial superconductor, and extensive transport studies have been carried out to unravel the orientation dependence [9,10,11,12,13,14,15,16]. To understand the microscopic origin of the interfacial superconductivity, it is imperative to develop alternative ways to tune the interfacial electron gas and superconductivity. In addition, it is also important to reveal the interfacial electronic states, which is often challenging for surface-sensitive electron spectroscopic measurements, such as angle-resolved photoemission spectroscopy (ARPES) or scanning tunneling microscopy (STM).

Previously, the two-dimensional electronic states associated with the $KTaO_3$ surface were studied for *in-situ* cleaved surfaces using synchrotron radiation ARPES measurements, where the electron gases were created by photon illumination [17,18,19,20]. Recently, the interfacial electron gases of $KTaO_3$ were investigated for the $LaAlO_3/KTaO_3$ and $Al/KTaO_3$ interfaces [21,22,23,24,25]. While dispersive quasiparticle bands were observed previously [17,18,19,20,21,22,23,24,25], superconductivity was only reported in Ref. [24,25]. In particular, *ex-situ* soft x-ray ARPES measurements on superconducting $LaAlO_3/KTaO_3$ interfaces revealed satellite bands that could be attributed to the electron-phonon coupling [25]. Here in this paper, we report a different approach to tune and investigate the $KTaO_3$–based interfacial electron gas by depositing pure rare-earth (RE, being La, Ce and Eu) metals on $KTaO_3$ substrates using molecular beam epitaxy (MBE). The strong chemical reactivity of RE metals allows for easy electron doping at the $KTaO_3$ interface, and the flexible choice of RE metals with different numbers of 4f electrons provides a multiplex knob to modify the electron gas and superconductivity. In addition, our combination with *in-situ* ARPES measurement provides direct spectroscopic insight on the microscopic nature of the interfacial electron gases. Our results imply that the direction-dependent quasiparticle dispersion, especially the appearance of satellite bands suggestive of interfacial electron-phonon coupling similar to [25], which might be important for understanding the orientation-dependent interfacial superconductivity.

## II. EXPERIMENTAL METHODS

The interfacial electron gases at the RE/$KTaO_3$ interfaces were created using MBE without delivering oxygen. Before the deposition of RE metals, the $KTaO_3$ substrates [with (111), (110) and (001) orientations] were degassed at 300℃ for 30 mins in ultrahigh vacuum (with a base pressure < $2x10^{-10}$ mbar), resulting in sharp high-energy reflection electron diffraction (RHEED) patterns. The substrates were hold at room temperature while depositing high purity Eu, Ce or La (> 99%, from Alfa Aesar). The evaporation of RE metals was achieved by effusion cells that were hold at 480℃ (Eu), 1450℃ (Ce) and 1225℃ (La), respectively, yielding an evaporation rate of 0.6 Å/min (Eu), 0.5 Å/min (Ce) and 1.5 Å/min (La), as verified by quartz crystal microbalance. We mention that depositions of Al and Eu metals at room temperature were used previously to generate two-dimensional electron gases at the oxide surfaces [26,27,28].

All the ARPES data were taken in a Helium-lamp ARPES system at Zhejiang University, which is connected under ultrahigh vacuum (UHV) to the MBE system. ARPES measurements were performed immediately after the deposition of RE metals. A five-axis manipulator cooled by a closed-cycle helium refrigerator was employed for the ARPES measurements. The photon source was a VUV-5k Helium lamp coupled to a grating monochromator for selecting either the He-I (21.2 eV) or He-II (40.8 eV) lines. The base pressure of the ARPES system was $7x10^{-11}$ mbar, which increased to $1.4x10^{-10}$ mbar during the Helium lamp operation. Most of the ARPES data were taken with He-I photons, except the core level scans [shown in Figs. 3(a)-3(c)], which were taken with He-II photons. Due to the low photoemission counts of the bands near Fermi energy ($E_F$), we used pass energy of 10 eV and 1.6 mm entrance slit, which yielded a typical energy resolution of ~20 meV and momentum resolution of ~0.01 $Å^{-1}$. ARPES measurements were performed at different coverages of RE metals. The interfacial electron gas begins to develop at very low coverage and becomes somewhat saturated around 1 Å RE coverage (corresponding to ~ 0.3 monolayer).

Although *in-situ* ARPES measurements were performed mostly with ~1 Å coverage of RE, for transport measurements, a very thin (~3 nm) layer of RE metals was deposited afterwards to protect the interfacial electron gas. A 10 nm amorphous Si layer was further deposited on top as the capping layer to protect against the oxidation in air. Note that a 3 nm layer of RE metals is necessary for protecting the $KTaO_3$ interface due to the intermixing between RE layer and Si capping layer. The *ex-situ* transport measurements were performed in a Quantum-design Physical Property Measurement System (PPMS), using the standard four-terminal method. The terminal leads were made via wire bond to allow access of the interfacial electron gases.

To understand the electronic doping from RE metal, electronic structure calculations for bulk $KTaO_3$ were performed using density functional theory (DFT) and a plane-wave basis projected augmented wave method, as implemented in the Vienna ab initio simulation package (VASP) [29,30]. The Perdew, Burke and Ernzerhof (PBE) generalized gradient approximation for exchange correlation potential was used for the DFT calculation [31]. Spin-orbital coupling (SOC) was taken into account in all DFT calculations. An energy cutoff of 600 eV and a 12×12×12 gamma-centered k-mesh were employed. A Fermi level shift of 2.09 eV compared to bulk $KTaO_3$ was employed to simulate the electron doping from the RE metals, which is ~0.17 $e^-$ per surface unit cell for Eu/$KTaO_3$(111) (considering the spin degeneracy), close to the experimental value. Such a large energy shift from a small doping is due to the large intrinsic band gap of $KTaO_3$.

## III. RESULTS AND DISCUSSIONS

### A. SAMPLE GROWTH AND CHARACTERIZATION

A schematic drawing of the experimental approach and the $KTaO_3$ crystal structure are shown in Figs. 1(a) and 1(b). The growth and formation of the RE/$KTaO_3$ interfaces were monitored by *in-situ* RHEED, as shown in Fig. 1(c) (additional data in supplementary Fig. S1). During the first 1-2 Å deposition of the RE metals, the RHEED patterns remain sharp, which are similar to the bare $KTaO_3$ substrate, indicating well-ordered interfacial structures. As we shall see below, photoemission measurements show that the interfacial electron gas is already formed at this stage. Upon further deposition, the RHEED patterns become ring-like with three-dimensional characters, implying that the subsequent deposition leads to amorphous RE films.

To probe the interfacial superconductivity, we first deposit ~3 nm RE metals, followed by ~10nm amorphous Si. This configuration serves for three purposes: 1) to allow full development of the interfacial electron gases; 2) to protect against the heating effect during the deposition of Si and intermixing with the Si capping layer; 3) to avoid sample oxidation in air because of the extremely high chemical reactivity of rare-earth metals; see also Fig. 1(a). The zero-field transport results are summarized in Fig. 2(a). The Eu/$KTaO_3$(111) interface shows superconductivity with the highest $T_c$ of ~2.2 K (here we define half of normal-state resistivity as $T_c$), while the $T_c$ of the Eu/$KTaO_3$(110) interface is ~1.0 K. The Eu/$KTaO_3$(001) interface reveals no sign of superconductivity at the lowest measurement temperature (0.5 K). Such orientation-dependent interfacial superconductivity is similar to previous results in both EuO/$KTaO_3$ and $LaAlO_3$/$KTaO_3$ interfaces [6,7,8]. The superconducting transition temperature of the La/$KTaO_3$(111) interface is lower than Eu/$KTaO_3$(111), with $T_c$ ~ 0.8 K, and the La/$KTaO_3$(001) interface is not superconducting down to 0.5 K, same as Eu/$KTaO_3$(001). By contrast, no sign of superconductivity can be observed for the Ce/$KTaO_3$ interfaces with all crystalline orientations. Since La, Ce and Eu shares similar chemical reactivity and differs mainly in the number of 4f electrons (0 for La, 1 for Ce and 7 for Eu), this non-monotonic evolution of $T_c$ implies that the 4f-electron filling can play a nontrivial role in tuning the interfacial superconductivity. We emphasize that bulk Ce and Eu are magnetic and not superconducting, while bulk La and LaO are both superconducting with $T_c$ ~ 6 K. Therefore, the superconductivity in Eu/$KTaO_3$ must be from the interfacial electron gases. On the other hand, the strong orientation dependence in the superconductivity of the La/$KTaO_3$ samples implies that the observed superconductivity in La/$KTaO_3$(111) is most likely from the interface as well, instead of the thin La (or LaO) layer on top.

To further understand the interfacial superconductivity, we applied a magnetic field perpendicular to the sample surface and studied the field dependence of the superconductivity, as summarized in Figs. 2(b)-2(e). It is clear from the data that the superconductivity can be continuously suppressed by the magnetic field. The extracted temperature-dependent upper critical field ($H_c$) curves are presented in Fig. 2(b), where linear fitting is used to analyze the critical fields as a function of temperature, based on the standard Ginzburg-Landau theory. Our fitting results yield that the critical fields for the Eu/$KTaO_3$(111), Eu/$KTaO_3$(110) and La/$KTaO_3$(111) interfaces (at zero temperature) are 2 T, 0.7 T and 0.5 T, respectively. Note that the out-of-plane critical field of Eu/$KTaO_3$(111) is similar to previous study on EuO/$KTaO_3$(111) interfaces [6], apparently below the Pauli limit. The similar relationship between the critical field and temperature, despite the differences in the values of $T_c$, implies that the superconductivity in these interfaces shares similar origin.

Previous transport studies have found direct correlation between $T_c$ and interfacial carrier density ($n_{2D}$) in $KTaO_3$-based interfaces [6,10,12,13,14]. In Fig. S3, the results of Hall resistance measurements for Eu/$KTaO_3$(111), Eu/$KTaO_3$(110), La/$KTaO_3$(111) and Ce/$KTaO_3$(111) samples are displayed, where the extracted the carrier densities are $2.06\times10^{14}$, $1.42\times10^{14}$, $5.06\times10^{14}$ and $1.67\times10^{15}$ cm$^{-2}$, respectively. Note that the $T_c$ values of our Eu/$KTaO_3$(111) and (110) samples are close to the highest reported values in literatures, implying a saturated interfacial carrier density consistent with the Hall measurements. On the other hand, both La/$KTaO_3$(111) and Ce/$KTaO_3$(111) have higher carrier density despite suppressed superconductivity. Thus, the element-dependent behavior should be explained by some other factors, e.g., the possible magnetism from Ce ions. The Ce/$KTaO_3$(111) curve is not completely linear [see Fig. 3(d)], implying the possible existence of magnetism or multiple bands near $E_F$. Based on the overall comparison between different RE elements on $KTaO_3$(111) interfaces, $T_c$ values of RE/$KTaO_3$ are not solely dependent on the measured $n_{2D}$. We mention that due to the presence of 3 nm RE layer, the extracted carrier density in Fig. S3 might not accurately reflect the contribution from the interface.

## B. SHALLOW CORE LEVELS AND VALENCE BANDS

To investigate the non-monotonic evolution of $T_c$ with RE metals, it is important to first understand the charge transfer and valence change at the interface. Therefore, we performed *in-situ* photoemission measurements of the core levels as a function of RE metal coverages, as shown in Figs. 3(a)-3(c). The starting $KTaO_3$(111) substrate is highly insulating with no electronic states near $E_F$, consistent with its large band gap. The broad peak from -5 eV to -10 eV can be attributed to the O 2p states. With La deposition, the spectral intensity below $E_F$ (0 to -5 eV) begins to grow, while the intensity of the O 2p bands is gradually reduced [Fig. 3(a)]. For the Ce/$KTaO_3$(111) case, an additional peak at ~-2.5 eV emerges [Fig. 3(b)], corresponding to the $Ce^{3+}$ 4f state, which is often seen in Ce-based Kondo lattice systems [32,33,34]. By contrast, the spectral evolution for the Eu/$KTaO_3$(111) interface is quite different from the former cases [Fig. 3(c)]: the spectral intensity between -5 eV and -10 eV does not show any decrease within the first few Å deposition, in sharp contrast with the La/$KTaO_3$(111) and Ce/$KTaO_3$(111) cases. In fact, a zoom-in view [inset in Fig. 3(c)] indicates that the intensity increases slightly for coverages below 2 Å. Since this energy window coincides with the 4f multiplets associated with $Eu^{3+}$ ions [35,36], it is natural to think that appreciable $Eu^{3+}$ component is present near the interface, due to charge transfer from the $KTaO_3$ substrate. Note that bulk Eu is normally divalent with seven 4f electrons, i.e., $Eu^{2+}$ with $4f^7$ configuration, as verified by the gradual increase of the peak at ~-2 eV that corresponds to the $Eu^{2+}$ 4f peak ($4f^6$). The core level scans for the Eu/$KTaO_3$(111) interface imply that the interfacial Eu has a non-negligible $Eu^{3+}$ component and is likely mixed-valent with simultaneous contributions from both $Eu^{3+}$ and $Eu^{2+}$ (since the $Eu^{2+}$ 4f peak also appears right after Eu deposition). This indicates the possible formation of both $Eu_2O_3$ and EuO at the interface.

The observed valence states of the RE metals provide the basis to understand the non-monotonic evolution of $T_c$. For the Ce/$KTaO_3$(111) case, since trivalent Ce with one 4f electron hosts local magnetic moment, it can be very detrimental to superconductivity if the superconductivity is driven by electron-phonon coupling within the Bardeen-Cooper-Schrieffer framework [37,38,39]. For the Eu/$KTaO_3$(111) case, although the bulk Eu ($4f^7$) is magnetic with large ordered moment, the interfacial Eu is mix-valent with large contributions from both $Eu^{3+}$ ($4f^6$) and $Eu^{2+}$ ($4f^7$), which could host superconductivity as other mixed-valent Eu-based superconductors [40,41].

The evolution of the photoemission intensity near $E_F$ is summarized in Fig. 3(d) for the Eu/$KTaO_3$(111) interface. For bare $KTaO_3$(111) substrate, no appreciable intensity can be observed. With only a small amount of Eu deposition, a small quasiparticle peak emerges near $E_F$, indicating the formation of interfacial electron gas. With further Eu deposition, the height of this peak does not increase significantly, except that the overall background becomes stronger due to the formation of amorphous Eu on top. Since the quasiparticle peak associated with the interfacial electron gas is best identified at the thickness of 0.5 Å (~ 0.15 monolayer), we focus on spectra taken at this coverage thereafter. We emphasize that 0.5 Å Eu (if becomes $Eu^{2+}$ and $Eu^{3+}$ at the interface) is able to provide sufficient electron density ($n_{2D} > 2*10^{14}/cm^2$), larger than the saturated value for $KTaO_3$(111) reported in ref [14] ($n_{2D} \sim 1*10^{14}/cm^2$, corresponding to $T_c$~2.2K). Moreover, the spectra obtained from the 0.5 Å layer show no qualitative difference compared to those from the 1-2 Å layers, further supporting that the 0.5 Å spectra capture the intrinsic characteristics of the interfacial superconductivity.

## C. QUASIPARTICLE DISPERSION

The quasiparticle dispersions of the interfacial electron gases for the Eu/$KTaO_3$(111), Eu/$KTaO_3$(110) and Eu/$KTaO_3$(001) interfaces are summarized in Fig. 4, where dispersions along two high-symmetry in-plane directions for the three cases are shown in Figs. 4(a)-4(f). For the Eu/$KTaO_3$(111) electron gas, the quasiparticle bands consist of a dispersive, light electron pocket [indicated by the green arrow in Fig. 4(d)] centered at $\bar{\Gamma}$, as well as a broad dispersionless feature (indicated by the light blue arrow, see Fig. S5 for detailed EDC plots) near $E_F$. The light electron band should be derived from the J=3/2 Ta 5d band in $KTaO_3$, which is pulled down below $E_F$ due to electron doping from Eu.

To understand the interfacial electronic structure, we also plot the calculated band structures of bulk $KTaO_3$ from DFT in Fig. 5. Although such bulk calculation cannot capture the detailed interfacial effects, it nevertheless illustrates a simple and crude way of understanding the interfacial electron gas from quantum confinement and electron doping. Specifically, the interfacial electronic state can be viewed as a quantum well state (QWS) in two-dimensional limit, which can be approximated by the bulk states with a specific $k_z$ value determined by the Bohr-Sommerfeld quantization condition [42,43]. The electron doping from Eu can be further simulated by an energy shift of the calculated bands. From the calculations, the observed light electron band should contain mixed contributions from all the $d_{xy}$, $d_{yz}$ and $d_{xz}$ orbitals, which are degenerate at Γ point and show minor anisotropy for the (111) surface. The experimental dispersions along two orthogonal in-plane directions are indeed similar [compare Fig. 4(a) and Fig. 4(d)], in reasonable agreement with DFT calculations. On the other hand, the dispersionless feature near $E_F$ cannot be easily explained by this simple DFT calculation. One possible reason is that the detailed interfacial effects require additional experimental inputs and more sophisticated theoretical approaches (see, e.g., [17,18,19,20,21,22,23]), which is beyond a simple simulation as shown in Fig. 5. In addition, the discrepancy might also originate from enhanced effective mass due to a combined effect of strong electron-phonon coupling (EPC) and disorder scattering (see below), or alternatively due to contributions from the RE 4f electrons. We admit that simple electron doping cannot fully explain the observed spectra. The factors missed by bulk calculations may be related to the origin of the orientation-dependent interfacial superconductivity. Since our paper is mainly an experimental work, more sophisticated calculations and comparison with experiments are beyond the scope of the current paper.

For the Eu/$KTaO_3$(110) interface, the dispersionless band near $E_F$ remains strong [light blue arrow in Fig. 4(e)], but the dispersive light bands (green arrow) become much weaker and difficult to identify clearly. For the non-superconducting Eu/$KTaO_3$(001) interface, no well-defined quasiparticle peak can be identified near $E_F$, implying possibly an incoherent electron gas, although it could also be attributed to the suppressed spectral intensity under 21.2 eV photons, which is not unusual for the interfacial electron gases, e.g., in [20], due to variation of photoemission matrix elements. Nevertheless, the different electronic structure between the (111)/(110) interfaces and the (001) interface provides a basic framework for understanding the orientation-dependent superconductivity. Note that the experimental quasiparticle bands cannot be simply explained by electron doping of bulk $KTaO_3$, based on a direct comparison between Fig. 4 and Fig. 5. Such a difference could be attributed to the simplified treatment based on bulk $KTaO_3$ calculations, and the nontrivial interfacial effects.

## D. DOUBLE-PEAK STRUCTURE FROM EPC

It is interesting to note that both the experimental EDCs for the Eu/$KTaO_3$(111) and Eu/$KTaO_3$(110) interfaces feature a double-peak structure [Figs. 4(g) and 4(h)], with a main band near $E_F$ and a shoulder-like structure at a lower energy. The energy separation between the main peak and its shoulder is around 100 meV, which is close to the energy of one longitudinal optical phonon (LO4) of the $KTaO_3$ substrate at Γ point [44]. Such a double-peak structure has also been observed in the Ce/$KTaO_3$(111) and La/$KTaO_3$(111) interfaces (see Fig. 6), as well as in a recent soft x-ray ARPES study on the $LaAlO_3$/$KTaO_3$ interfaces [25], implying a similar origin. Such spectral feature could be explained by formation of replica bands as a result of electron-phonon coupling (EPC), as often observed in other low-dimensional systems with strong EPC [45,46,47]. Other possible explanations are that the double-peak feature may arise from the light and heavy bands, or the formation of multiple QWSs due to surface band bending [42], although such scenarios are difficult to explain the similar energy separation among different interfaces (see below).

To understand the origin of the double-peak features, we performed detailed fittings of the experimental EDCs based on the aforementioned scenario of EPC, as shown in Figs. 4(g) and 4(h) (more details are included in supplementary materials). We employed two Gaussian peaks to account for the main band and the first-order satellite band: we ignore higher-order effects since they are not obvious from the raw data. The full width at half maximum (FWHM) of these two peaks is constrained to be identical, to minimize the fitting parameters. A second-order polynomial is used to mimic the secondary electron background, which is obtained through fitting EDCs below -0.25 eV. The Fermi-Dirac distribution and the finite energy resolution are both considered. The fittings can reproduce the experimental EDCs quite well [red curves in Figs. 4(g) and 4(h)].

Fitting results for different RE/ $KTaO_3$ interfaces are summarized in Table 1, which also include results from different samples of the same type (see supplementary Fig. S2). It is intriguing to see that for all the interfaces, where the double-peak structure is obvious in their EDCs, the fittings consistently give an energy separation around 106 meV, close to the LO4 phonon energy of $KTaO_3$ at Γ point (102.4 meV) [44]. The consistency among different $KTaO_3$ interfaces implies that the strong EPC might be the origin of the double-peak structure.

Referring to the standard theory of EPC, the intensity ratio between the replica (or satellite) band and the main band can be directly related to the strength of EPC. Detailed comparison between different samples implies that the ratio seems to be dependent mainly on the crystalline orientation and insensitive on the type of RE metals, with a typical value of ~0.48 for (111) interfaces and ~0.40 for (110) interfaces, respectively. Such a dependence suggests that, if the double-peak structure was indeed from interfacial EPC, the EPC for the (111) interface could be stronger than the (110) interface. Nevertheless, since reliable extraction of EPC parameters often requires high-quality experimental data with well-defined satellite peaks and small background, future experimental studies with improved spectral quality are required to verify such an orientation dependence.

## E. DISCUSSION

Our combination of MBE growth and *in-situ* ARPES measurements allows us to investigate the intrinsic properties of the interfacial electron gases. Note that for photon energies used in the current study, we do not observe any signature of the photon-induced metallic states, which were observed previously for synchrotron-based ARPES measurements on cleaved $KTaO_3$ samples [17,18,19]. In addition, our complementary transport measurements of superconductivity allow us to make direct connection with the electronic structures, which can be important for a consistent understanding of the $KTaO_3$-based superconductivity.

The dispersionless features near $E_F$ [Figs. 4(a), 4(b), 4(d) and 4(e)] cannot be easily explained by DFT calculations. One possible explanation for these signals is the formation of polarons as a result of strong EPC and disorder scattering, which conspire to give rise to quasiparticle bands with large effective mass. Such phenomena have been observed in other systems, such as $La_{2-x}Sr_xTiO_4$, and can be in play in the current case as well. Small contributions from 4f electrons are also possible, since the dispersionless features for the Ce/$KTaO_3$ and Eu/$KTaO_3$ interfaces are slightly stronger than that of the La/$KTaO_3$ interface (along the (111) orientation). Nevertheless, future studies are still needed to unravel the detailed origin of the dispersionless features and their connection with the superconductivity.

While the correlation between the strength of EPC (inferred from the intensity ratio between the main and satellite band in Table 1) and superconductivity provides a possible explanation for the orientation-dependent superconductivity, as reported in [25], the non-monotonic evolution of $T_c$ with the RE element reflects the sensitivity of superconductivity with the 4f-electron filling. Comparing with no 4f electron La/$KTaO_3$ system, the existence of 4f electrons (in Ce and Eu) could either suppress superconductivity (Ce, likely through magnetism) or enhance it (Eu, possibly by mixed valency of Eu). Note that the mixed valency in Eu is known to suppress its magnetism and might give way to superconductivity [40,41].

According to Table 1, the intensity ratio between the replica band and main band is close to 0.48 and 0.40 for the (111) and (110) interfaces, respectively. Below we propose a possible explanation for such orientation dependence, based on a simple geometric consideration inspired by Refs. [48,49]: Ignoring higher order effects, the intensity ratio between the replica band and the main band is approximately proportional to the strength of EPC ($\lambda$) [49]. Since the EPC here arises from coupling with the LO4 phonon of $KTaO_3$ substrates near the Γ point, $\lambda$ is directly related to the magnitude of the vibrating electric dipole field associated with this mode. The corresponding term in Hamiltonian is proportional to $\langle\frac{\partial V_{EI}}{\partial \boldsymbol{\tau}}\cdot\boldsymbol{u}\rangle = \langle\frac{\partial V_{EI}}{\partial \tau_x}u_x\rangle + \langle\frac{\partial V_{EI}}{\partial \tau_y}u_y\rangle + \langle\frac{\partial V_{EI}}{\partial \tau_z}u_z\rangle$, where $V_{EI}$ represents the Coulomb interaction between interfacial electrons and ions, $\boldsymbol{\tau}$ is the relative position between the electron and ion, and $\mathbf{u}$ is the relative displacement of the ion, respectively (here we define z direction perpendicular to the interface). Considering the anisotropic dielectric constant as well as interfacial electrons on top, $\frac{\partial V_{EI}}{\partial \tau_z}$ should be much larger than $\frac{\partial V_{EI}}{\partial \tau_x}$ and $\frac{\partial V_{EI}}{\partial \tau_y}$. In addition, the interfacial symmetry breaking leads to an extra electric-dipole distribution along z direction, which also contribute to the dominance of the z component due to the nature of dipole-dipole interaction. For the LO4 phonon mode at the Γ point, the sum of charge displacement $\mathbf{u}$ is along the [111] direction [see Fig. 1(b)], mainly contributed by the lighter oxygen ions. Since the thickness of the interfacial electron gas can be a few nanometers [6,7,15], the magnitude of vibration should be summed by unit volume (instead of one surface layer). The corresponding $\lambda$ is proportional to the projection of total displacement along the surface normal, leading to a ratio of $\sqrt{3}:\sqrt{2}:1$ for (111), (110) and (001) interfaces, respectively. Note that the estimated ratio, $\sqrt{3}:\sqrt{2} \approx 1.22$, is in reasonable agreement with the experimental value of 0.48:0.40 in table 1, as well as the extracted ratio in $LaAlO_3/KTaO_3$ interfaces in [25]. While our results indicate the strong orientation dependence of EPC, other factors, such as carrier density and orbital degeneracy [14], should also be in play and jointly give rise to the orientation-dependent superconductivity.

## IV. CONCLUSION

In summary, we developed a method of growing superconducting $KTaO_3$ interfaces by depositing RE metals and revealed their interfacial electronic structures using *in-situ* ARPES measurements. We found that the superconductivity exhibits non-monotonic change with RE metals: superconductivity is strongest for Eu, weaker for La and absent for Ce. Such a large difference points to the nontrivial role of 4f electrons in tuning the interfacial electron gas and its superconductivity. Our *in-situ* photoemission measurements revealed the mixed-valent behavior for Eu at the interfaces, which might be important for its enhanced superconductivity. Our APRES results also uncovered distinct quasiparticle dispersions near $E_F$ for RE/$KTaO_3$ interfaces with different orientations, suggesting the pivotal role of the orientation-dependent electronic structure. Double-peak spectral features can be observed near $E_F$ for all the (111) and (110) interfaces, and their energy separations are close to the energy of $KTaO_3$'s LO4 phonon at the Γ point. Detailed analysis suggests that this double-peak structure could be related to the interfacial EPC, which may shed light on the origin of orientation dependent superconductivity in $KTaO_3$-based interfaces.

## ACKNOWLEDGMENTS

This work is supported by National Key R&D Program of the MOST of China (Grant No. 2022YFA1402200, No. 2023YFA1406303), Fundamental Research Funds for the Central Universities (2021FZZX001-03), Key R&D Program of Zhejiang Province, China (2021C01002), National Science Foundation of China (No. 12174331) and State Key Project of Zhejiang Province (No. LZ22A040007). We would like to thank Prof. Yanwu Xie, Prof. Donglai Feng, Prof. Haichao Xu, Prof. Andrés Syntander-Syro and Prof. Ming Shi for helpful discussions.

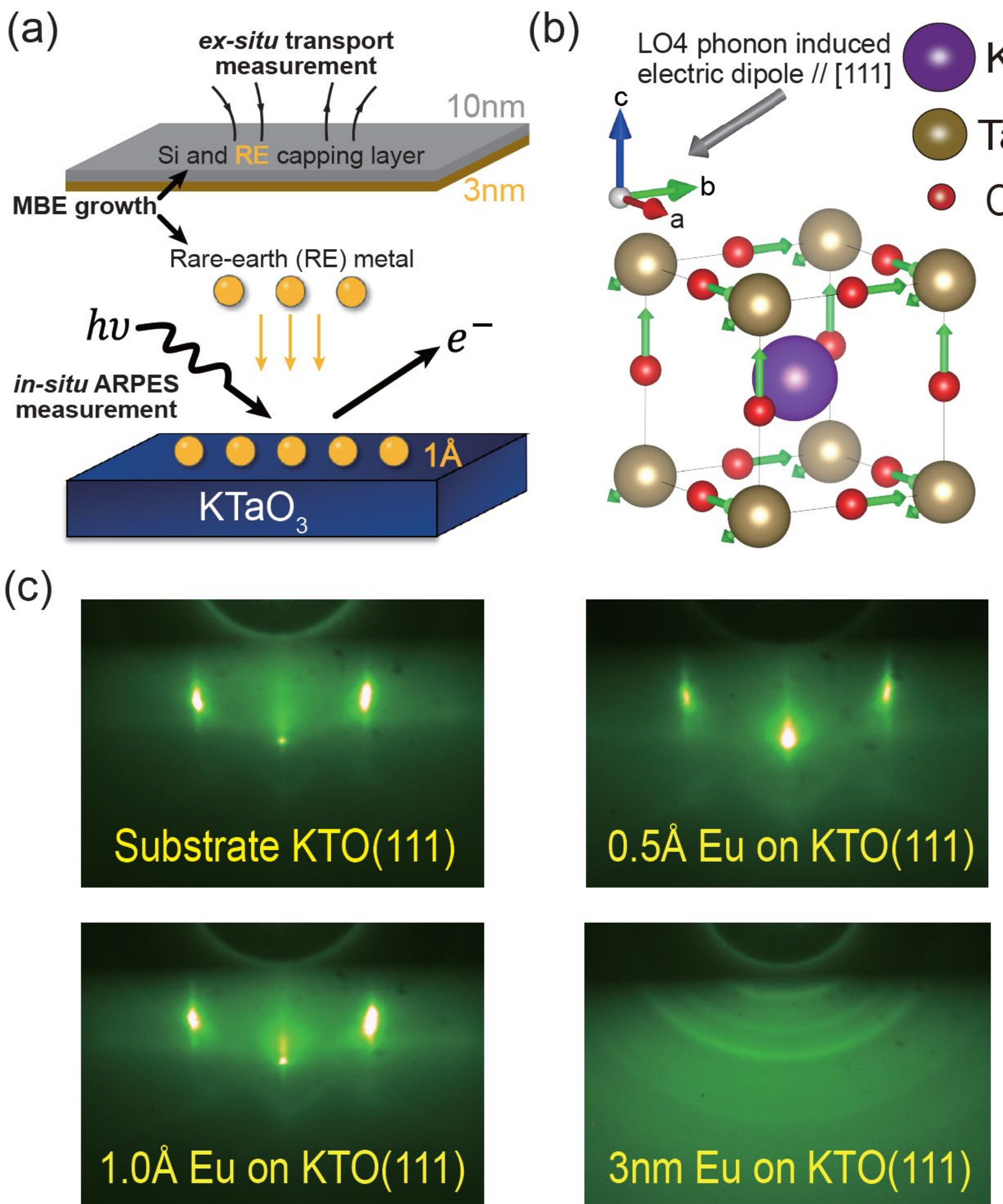

Figure 1. Formation of the RE/$KTaO_3$ interfaces. (a) A schematic view of how the interfacial electron gases are grown and probed in the paper. (b) Crystal structure of $KTaO_3$ substrates. The green arrows indicate the directions of ionic vibration corresponding to the LO4 phonon at Γ point. The grey arrow is the corresponding electric dipole parallel to [111]. (c) *In-situ* RHEED patterns during the different stages of Eu deposition for the Eu/$KTaO_3$(111) interface.

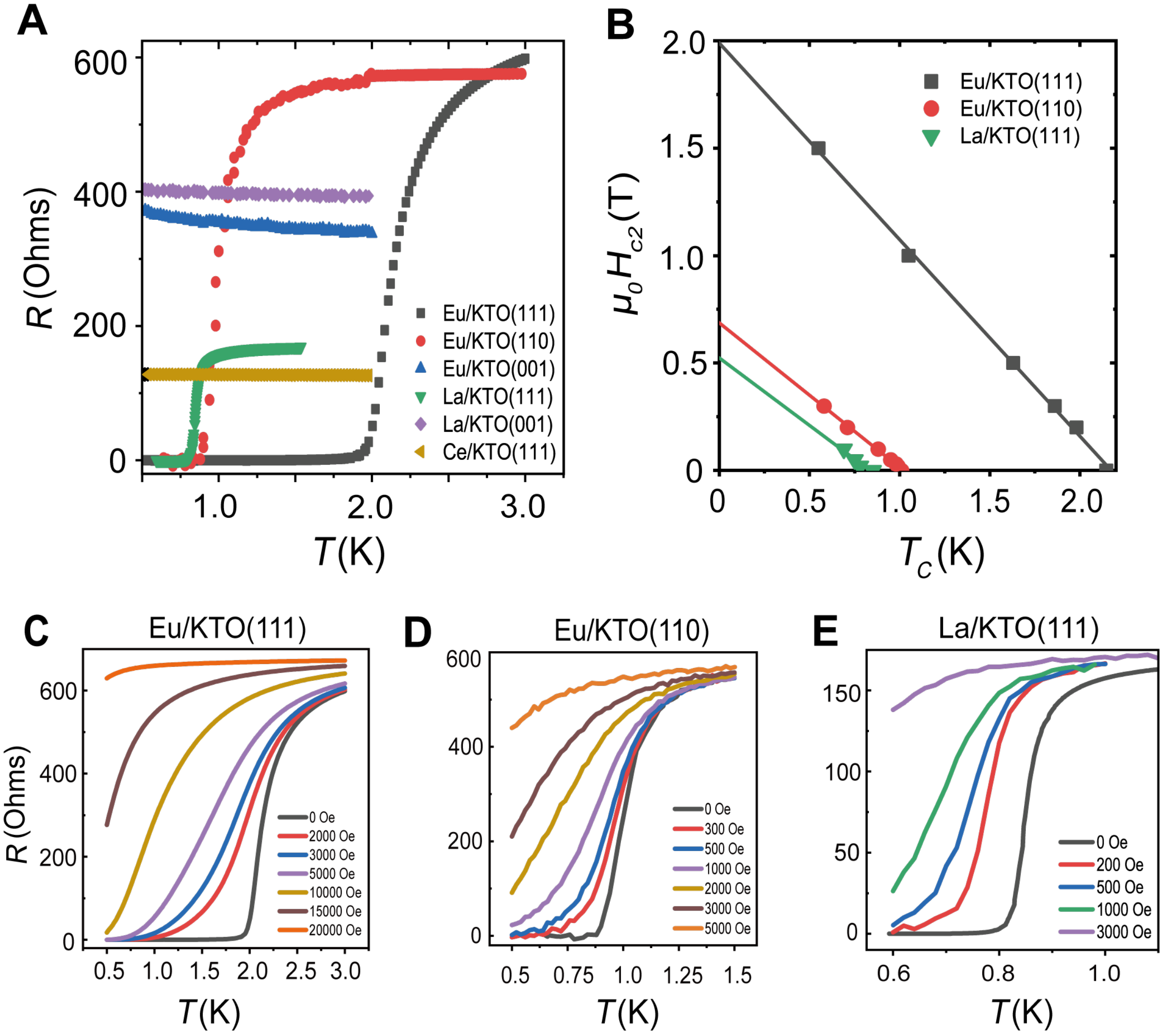

Figure 2. Superconductivity of the electron gases at the RE/$KTaO_3$ interfaces. (a) Resistance as a function of temperature for various RE/$KTaO_3$ interfaces along different crystalline orientations (see Fig. S2 for zoom-in view of the superconducting state). (b) Temperature dependence of out-of-plane upper critical fields for the superconducting Eu/$KTaO_3$(111), Eu/$KTaO_3$(110) and La/$KTaO_3$(111) interfaces. The solid lines are the fittings of the critical fields as discussed in the text. (c-e) Resistance as a function of temperature under different out-of-plane magnetic fields for superconducting Eu/$KTaO_3$(111) (c), Eu/$KTaO_3$(110) (d) and La/$KTaO_3$(111) (e) interfaces. The critical fields in (b) are extracted from the data in (c-e).

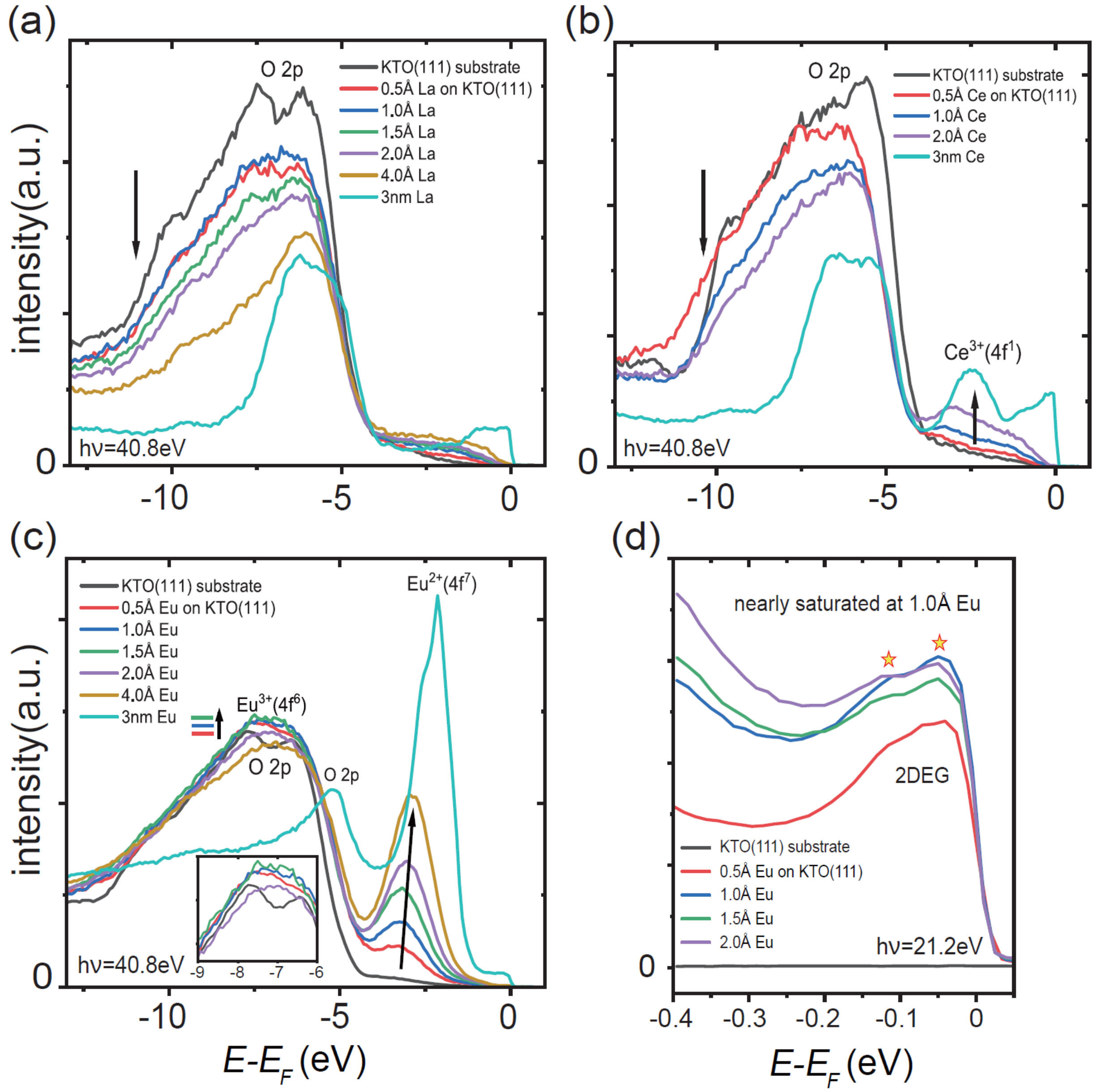

Figure 3. *In-situ* photoemission measurements of the shallow core levels and valence bands during the formation of the RE/$KTaO_3$(111) interfacial electron gases. (a-c) Momentum-integrated energy scans using He II (40.8 eV) photons at different RE coverages on $KTaO_3$(111): (a) for La, (b) for Ce and (c) for Eu. The inset in (c) is a zoom-in view for the spectral intensity from -9 eV to -6 eV. (d) The energy distribution curves (EDCs) near $E_F$ for the Eu/$KTaO_3$(111) interface taken with He I photons. The peak from the two-dimensional electron gas (2DEG) saturates near 1.0 Å, exhibiting a double-peak-like lineshape denoted by two stars. EDCs of thicker samples show slightly higher intensity near Fermi level, which is due to the dispersionless background, evident in higher tails at lower energy ($E-E_F \sim -0.4$ eV).

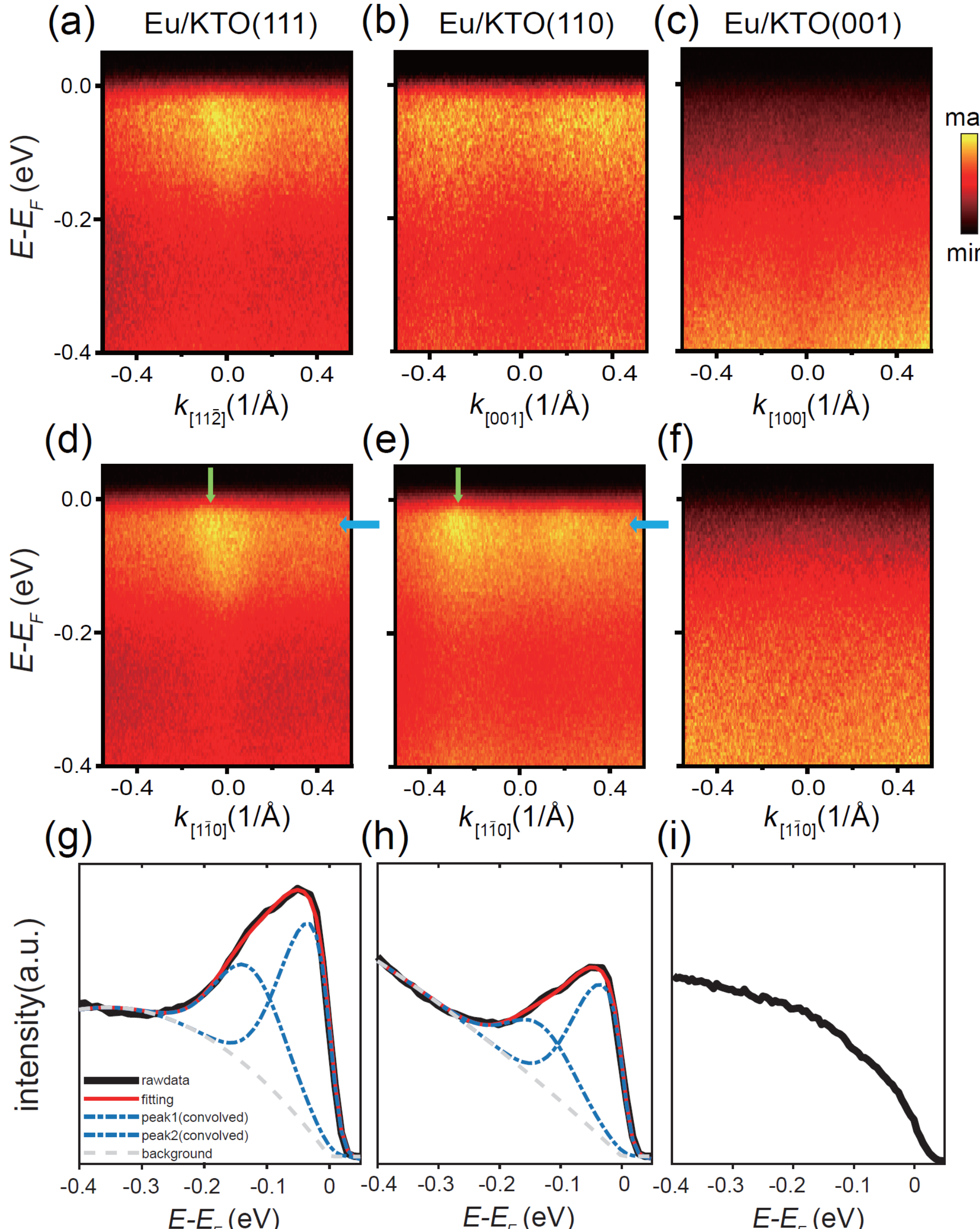

Figure 4. Quasiparticle dispersion near $E_F$ for the electron gases at the Eu/$KTaO_3$(111), Eu/$KTaO_3$(110) and Eu/$KTaO_3$(100) interfaces, obtained after deposition of 0.5 Å Eu. (a-f) Band dispersion for the Eu/$KTaO_3$ electron gases along two high-symmetry in-plane directions: (a, d) is for Eu/$KTaO_3$(111), (b, e) is for Eu/$KTaO_3$(110), (c, f) is for Eu/$KTaO_3$(001). The green and light blue arrows in (d, e) indicate dispersive and dispersionless bands, respectively. (g-i) are the corresponding EDCs (black curves) integrated between -0.25 and 0.25 $Å^{-1}$ for three interfaces. The fitting results (including two peaks and background) are also displayed for the (111) and (110) interfaces.

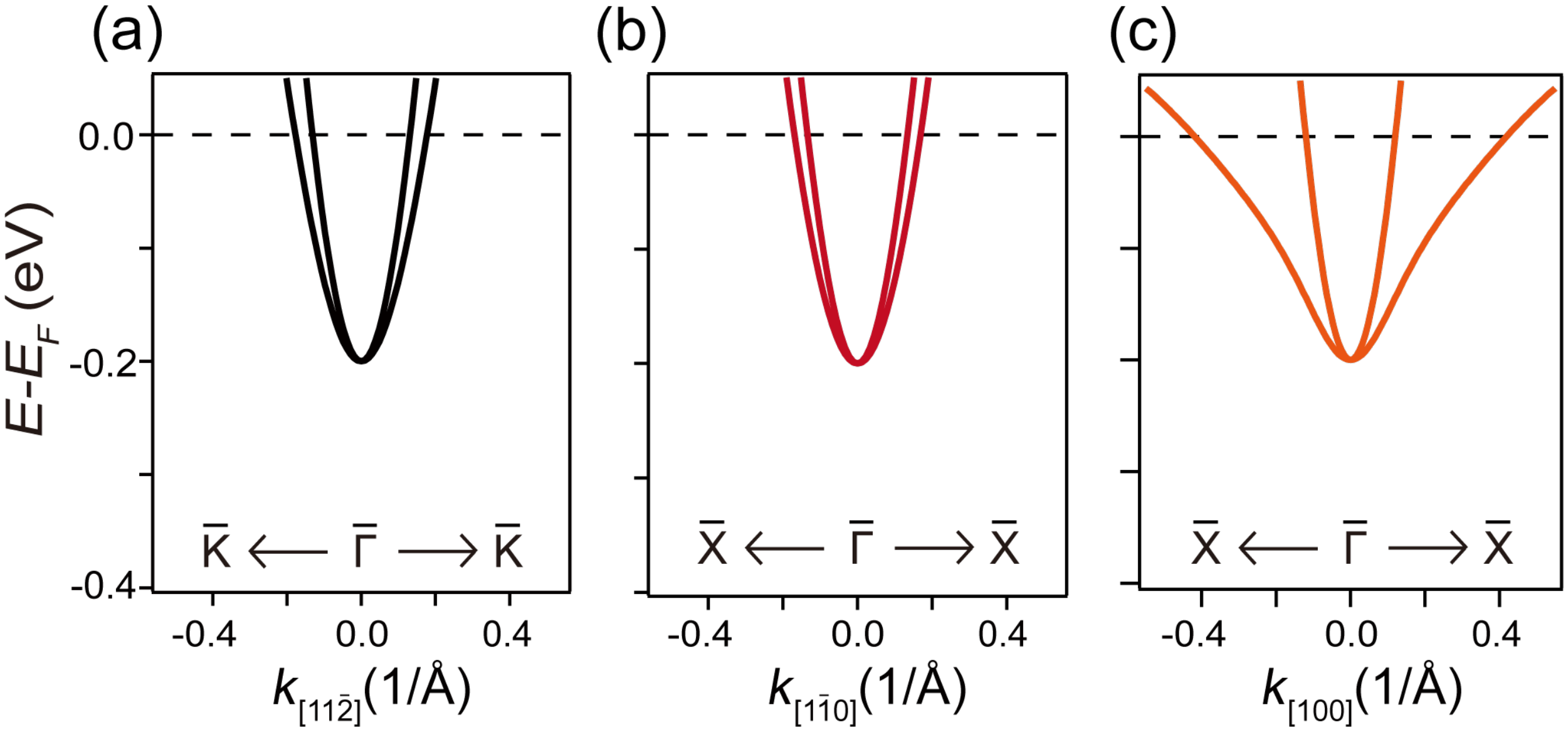

Figure 5. Calculated bulk band structures of $KTaO_3$ along(a) [11-2], (b) [1-10], (c) [100]. The Fermi level was shifted upward by 2.09 eV compared to undoped $KTaO_3$, whose Fermi level is at valence band maximum. This is to simulate the electron doping from the RE metal, which corresponds to ~0.17 $e^-$ per surface unit cell for (111) based on Luttinger theorem. The difference between these calculations and the experimental results in Figs. 5(a)-5(f) could be due to interfacial effects that are not included in the simple bulk calculations.

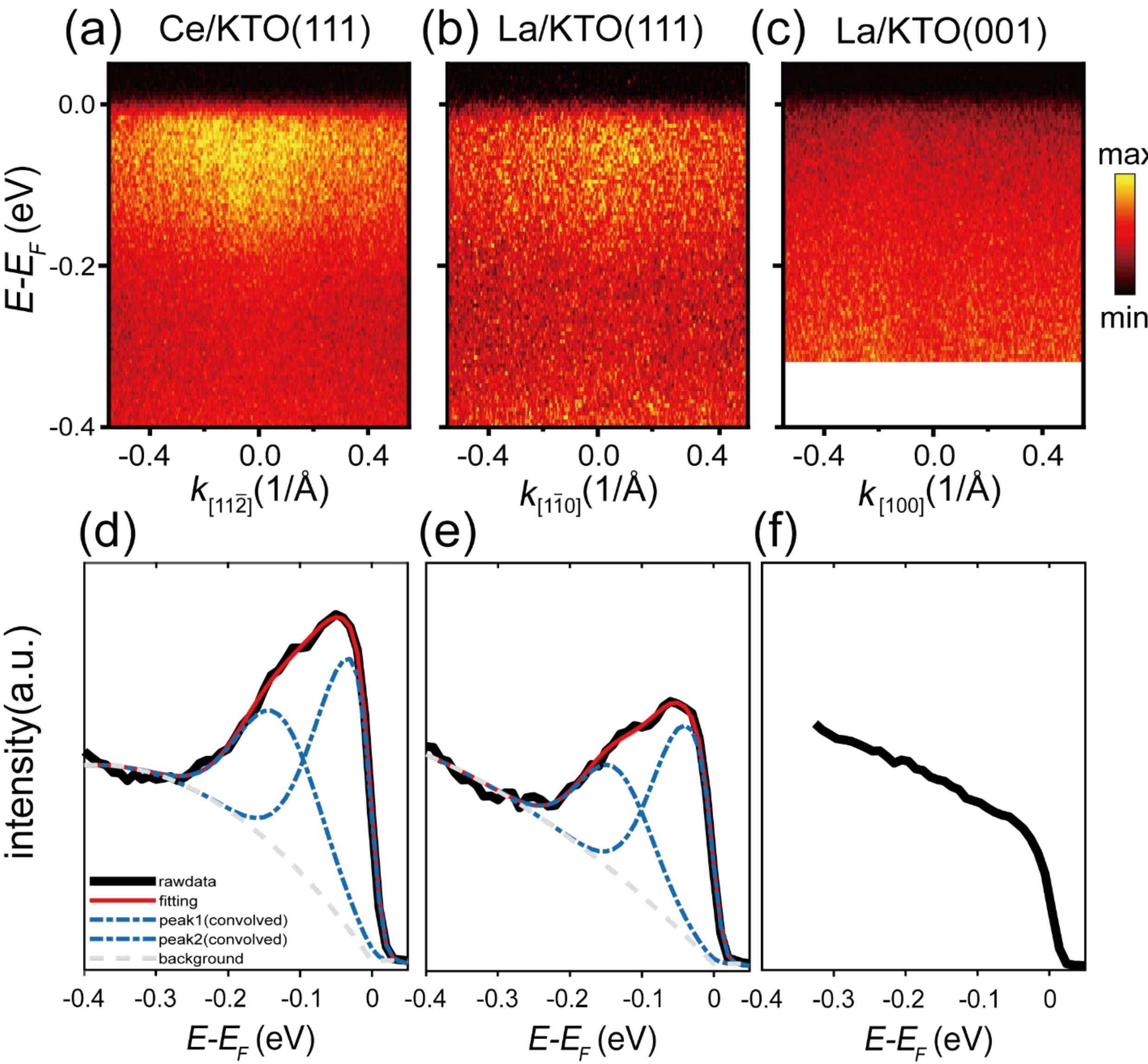

Figure 6. Quasiparticle dispersion near $E_F$ for the electron gases at the Ce/KTaO$_3$(111), La/KTaO$_3$(111) and La/KTaO$_3$(001) interfaces, obtained after deposition of 0.5 Å Ce or La. (a) is for Ce/KTaO$_3$(111), (b) is for La/KTaO$_3$(111) and (c) is for La/KTaO$_3$(001). (d-f) are the corresponding EDCs integrated between -0.25 and 0.25 Å$^{-1}$ for three interfaces. The fitting results are also indicated.

| Samples | Main band $E_b$ (meV) | Replica band $E_b$ (meV) | Energy separation (meV) | Intensity ratio $I_r/I_m$ | FWHM (meV) | $T_c$ (K) |
|---|---|---|---|---|---|---|
| Eu/KTa$O_3$(111) #1 | 22 | 125 | 103 | 0.48 | 76 | 2.2 K |
| Eu/KTa$O_3$(111) #2 | 25 | 131 | 106 | 0.46 | 78 | |
| Eu/KTa$O_3$(111) #3 | 11 | 120 | 109 | 0.49 | 90 | |
| Eu/KTa$O_3$(111) #4 | 21 | 127 | 106 | 0.48 | 78 | |
| Eu/KTa$O_3$(110) #1 | 25 | 134 | 109 | 0.41 | 77 | 1.0 K |
| Eu/KTa$O_3$(110) #2 | 27 | 138 | 111 | 0.40 | 77 | |
| Eu/KTa$O_3$(110) #3 | 21 | 131 | 110 | 0.41 | 77 | |
| Eu/KTa$O_3$(110) #4 | 22 | 128 | 106 | 0.39 | 78 | |
| Eu/KTa$O_3$(001) | - | - | - | - | - | No SC |
| La/KTa$O_3$(111) | 32 | 135 | 103 | 0.50 | 70 | 0.85 K |
| La/KTa$O_3$(001) | - | - | - | - | - | No SC |
| Ce/KTa$O_3$(111) | 19 | 128 | 109 | 0.49 | 80 | No SC |

Table 1. Summary of the fitting results for various RE/KTa$O_3$ interfaces. For Eu/KTa$O_3$(111) and Eu/KTa$O_3$(110) interfaces, results from four different samples (#1 to #4) are shown. Data from Eu/KTa$O_3$(111) #1, Eu/KTa$O_3$(110) #1, Eu/KTa$O_3$(001), La/KTa$O_3$(111), La/KTa$O_3$(001) and Ce/KTa$O_3$(001) are shown in Fig. 4 and 6, while data from the other samples are shown in supplementary Fig. S4. $E_b$ of the main and replica bands stand for their binding energies relative to $E_F$. The intensity ratio ($I_r/I_m$) indicates the ratio of the peak height between the replica band ($I_r$) and the main band ($I_m$).

# SUPPLEMENTARY MATERIAL

## I. Details of the EDC fitting with double peaks

Here we introduce the fitting process of EDC using two peaks. The function form of the fitting function is shown below:

$$EDC_{fit} = f_{BG} + \left[\left(I_{main} + I_{replica}\right) \cdot f_{FD}\right] * f_{\Delta E}$$

$$f_{BG} = \begin{cases} ax^2 + bx & if\ x < 0 \\ 0 & if\ x \geq 0 \end{cases},$$

$$I_{main} = a_1 e^{-\frac{(x-b_1)^2}{c^2}}, I_{replica} = a_2 e^{-\frac{(x-b_2)^2}{c^2}},$$

$$f_{FD} = \frac{1}{e^{\frac{x}{k_B T}} + 1}, f_{\Delta E} = e^{-\frac{x^2}{\Delta E^2/4ln2}},$$

where x is the energy ($E-E_F$); $a_1$, $a_2$ is the height of main band and the first replica band; $b_1$, $b_2$ is the peak position and c is standard deviation of both Gaussian peaks. Our ARPES were performed with T = 6 K, and the energy resolution $\Delta E$ is around 20 meV, based on Gold Edge Analysis at Fermi level.

We determine the background individually in the first place. A second-order polynomial is used to mimic the secondary electron background, which is obtained through fitting EDCs below -0.25 eV. We restrict that the background is zero above Fermi level.

Once the background is determined for each sample, we substitute it to the $EDC_{fit}$ and fit it to the raw-data. The rest part of fitting only needs five fitting parameters, which are $a_1$, $a_2$, $b_1$, $b_2$ and c. It is hard to reduce the number of fitting parameters further. And the only restriction we applied here is that both peaks have the same full width at half maximum (FWHM).

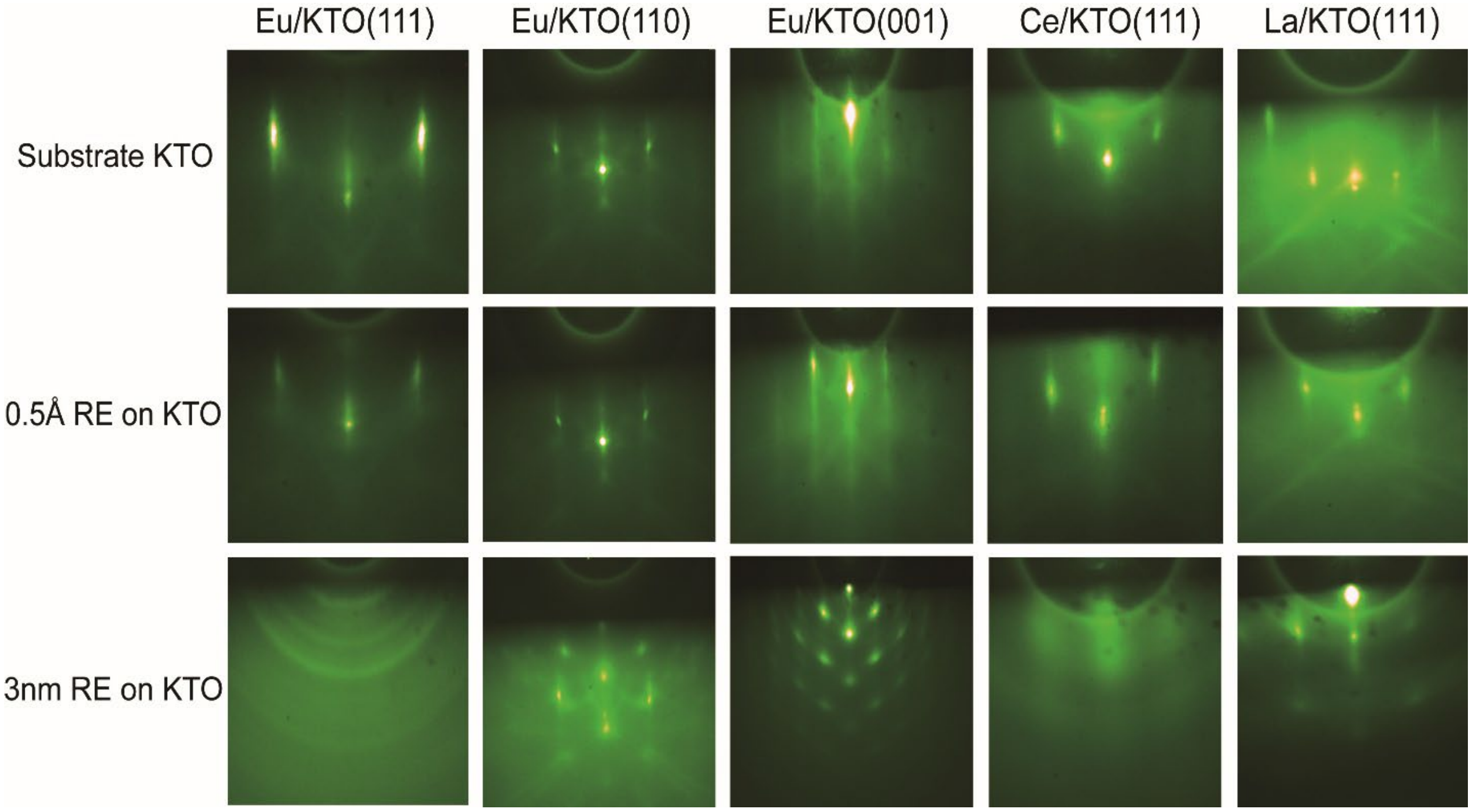

Figure S1. *In-situ* RHEED patterns during the different stages of Eu deposition for the Eu/$KTaO_3$ interfaces with different orientations. The RHEED patterns remain sharp during the first few Å's of Eu. Upon further deposition, the RHEED patterns become ring-like or spot-like with three-dimensional characters.

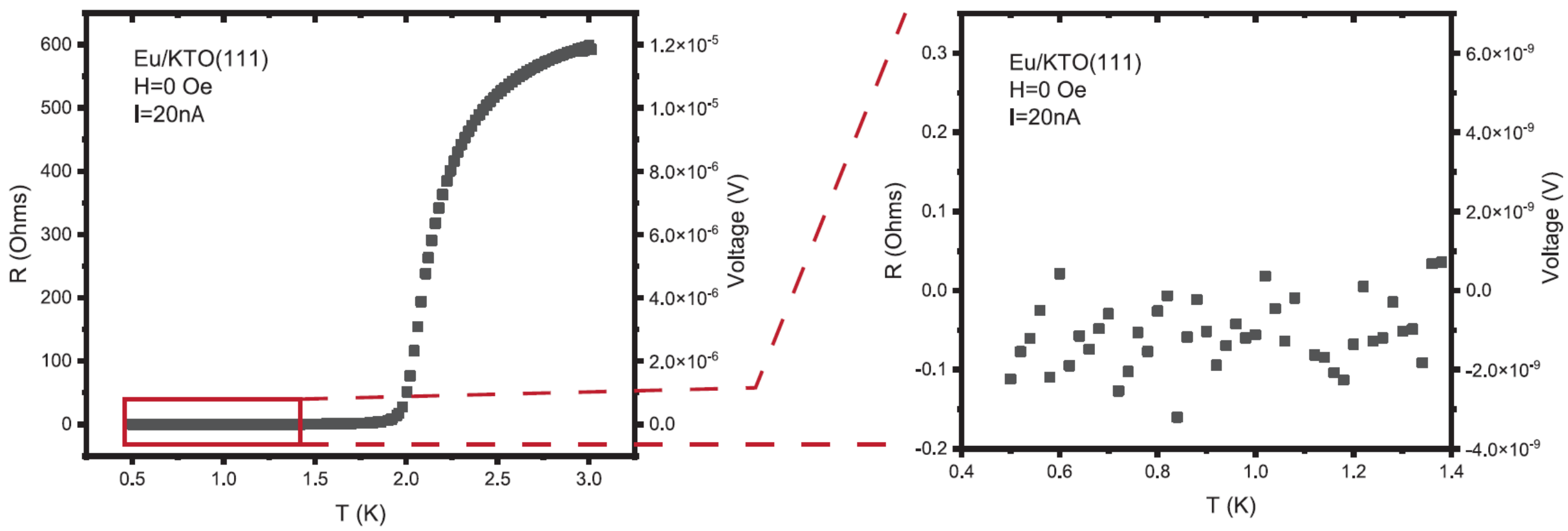

Figure S2. Zoom-in view of the resistance (and voltage) near the superconducting transition for a typical Eu/$KTaO_3$ (111) sample. Since a very small current is applied (to minimize the sample heating), the corresponding voltage is close to the detection limit of the voltmeter in PPMS.

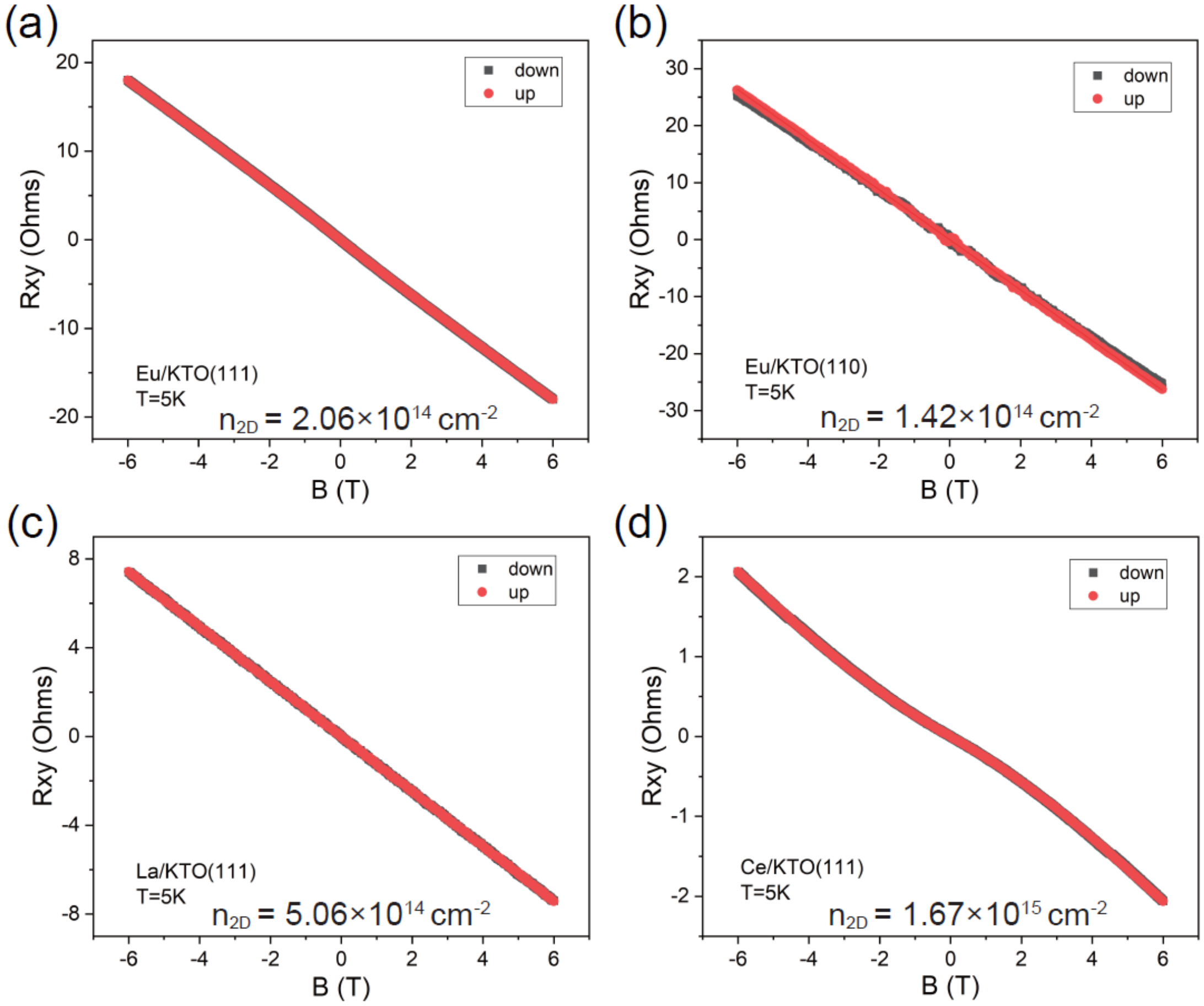

Figure S3. Hall effect measurements with external magnetic field perpendicular to the sample surface. The Hall signal and calculated interfacial carrier density of (a) Eu/$KTaO_3$(111), (b) Eu/$KTaO_3$(110), (c) La/$KTaO_3$(111) and (d) Ce/$KTaO_3$(111) interfaces are shown respectively. The red and black curves correspond to opposite magnetic field directions. Note that the 3nm RE layer inevitably contributes additional $n_{2D}$ in Hall measurements, so these values can only reflect properties of the whole device based on 10nm Si/ 3nm RE metal/ $KTaO_3$ configuration.

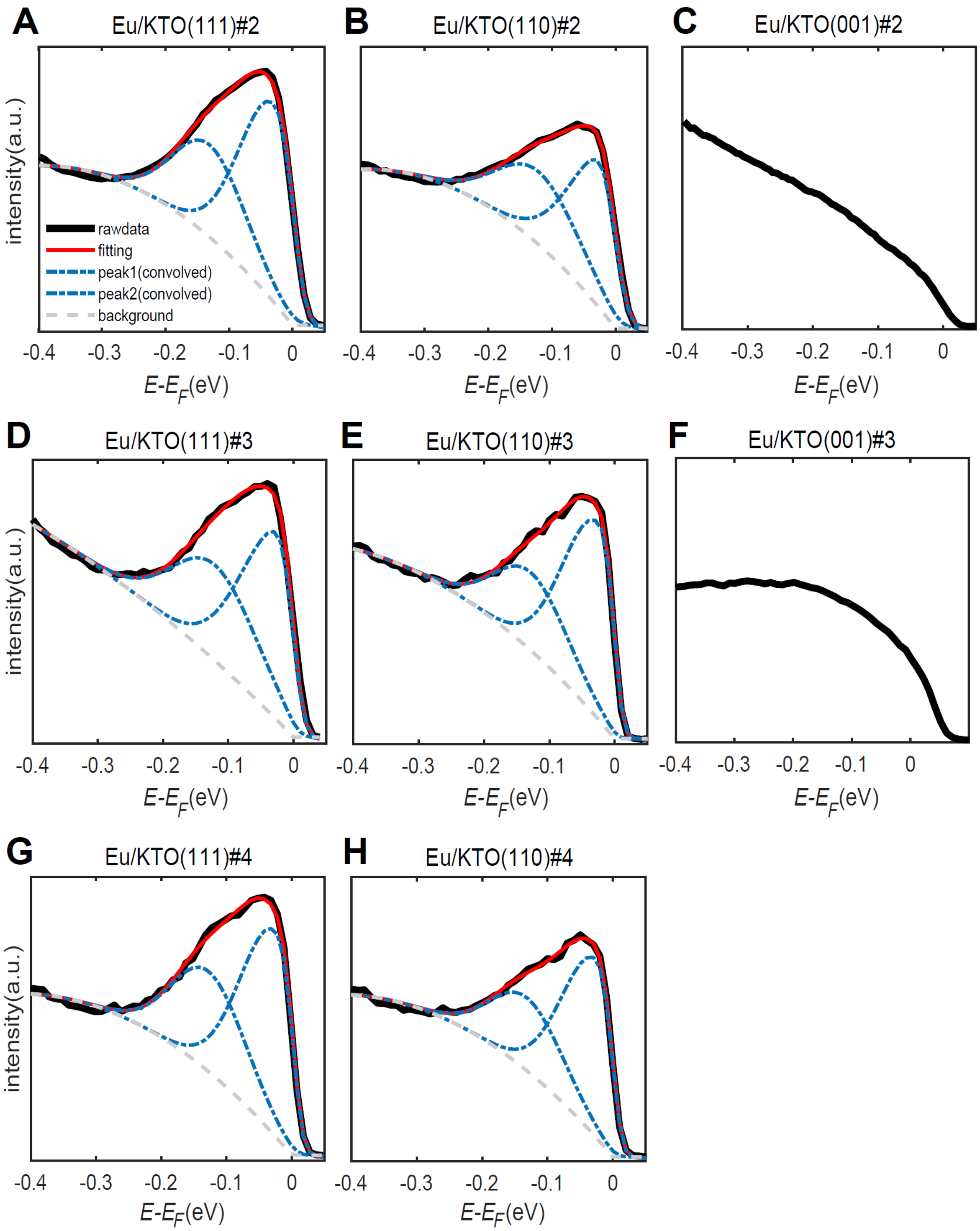

Figure S4. EDCs integrated between -0.25 and 0.25 Å$^{-1}$ for the electron gases at Eu/KTaO$_3$(111), Eu/KTaO$_3$(110) and Eu/KTaO$_3$(001) interfaces. Data from more samples (#2, #3, #4) for each orientation are shown. (a, d and g) are for Eu/KTaO$_3$(111), (b, e and h) are for Eu/KTaO$_3$(110), (c and f) are for Eu/KTaO$_3$(001).

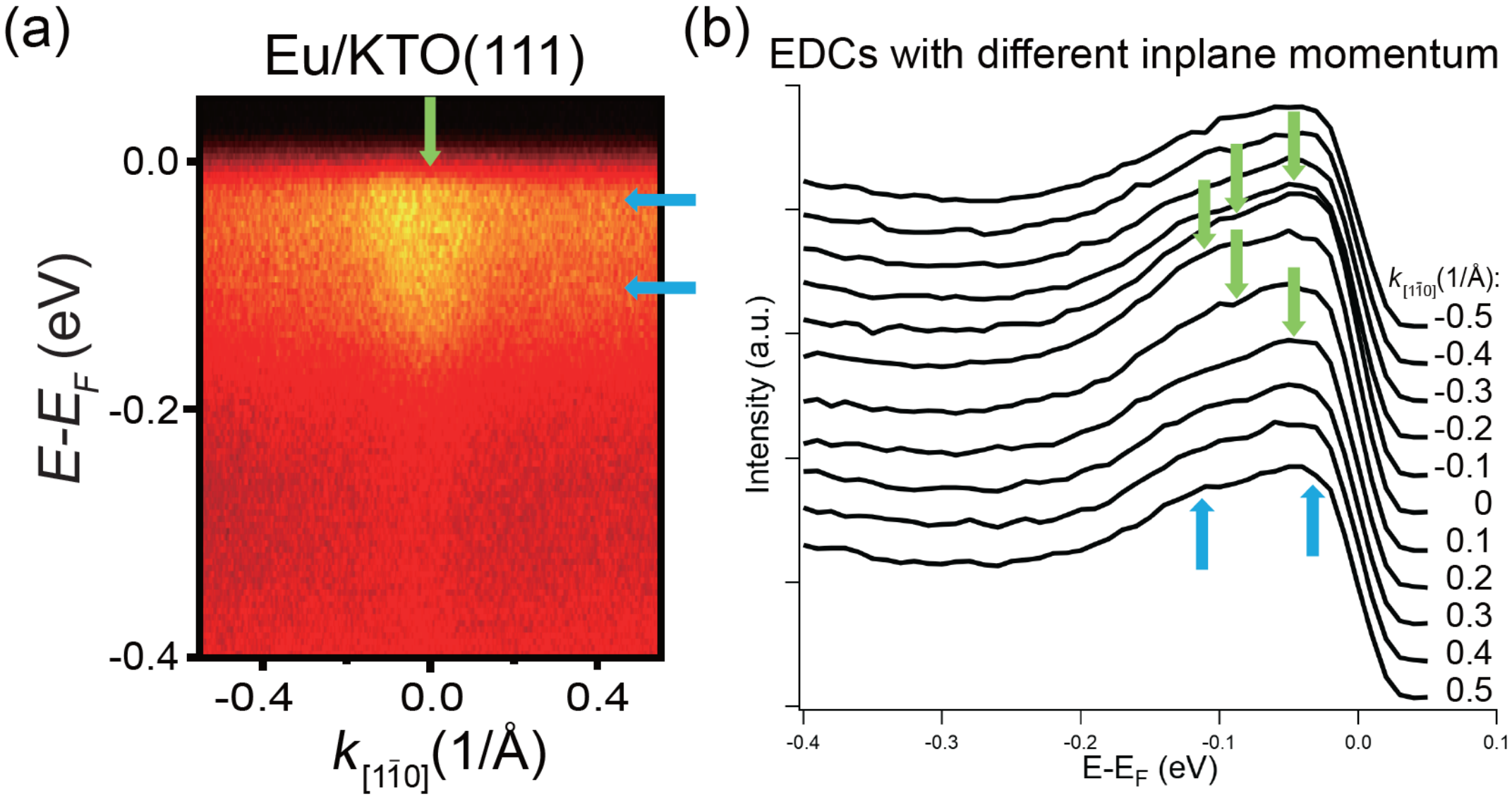

Figure S5. (a) Quasiparticle dispersion near $E_F$ of Eu/$KTaO_3$(111) interface, same as Fig. 5(d) in the manuscript. (b) EDC waterfall plot of (a) with different in-plane momenta. The green and light blue arrows indicate dispersive and dispersionless bands, respectively. It is clear from these data that the dispersionless band persists in the entire momentum range.